\documentclass[aps,prl,superscriptaddress,twocolumn]{revtex4-1}
\usepackage{graphicx}  
\usepackage{dcolumn}   
\usepackage{amsmath}
\usepackage{braket} 
\usepackage{lineno}
\usepackage{xcolor}
\usepackage{amssymb}

\usepackage{hyperref}
\hypersetup{colorlinks=true,
		    linkcolor=blue,
		    citecolor=blue,
		    allcolors=blue}

\usepackage{float}

\numberwithin{equation}{section}

\newcommand{\RNum}[1]{\uppercase\expandafter{\romannumeral #1\relax}}

\begin{document}

\title{Distinguishing a Mott Insulator from a Trivial Insulator with Atomic Adsorbates}

\author{Jinwon Lee}
	\affiliation{Center for Artificial Low Dimensional Electronic Systems, Institute for Basic Science (IBS), Pohang 37673, Republic of Korea}
	\affiliation{Department of Physics, Pohang University of Science and Technology, Pohang 37673, Republic of Korea}
	
\author{Kyung-Hwan Jin}
    \affiliation{Center for Artificial Low Dimensional Electronic Systems, Institute for Basic Science (IBS), Pohang 37673, Republic of Korea}
    
\author{Han Woong Yeom}
	\email{yeom@postech.ac.kr}
	\affiliation{Center for Artificial Low Dimensional Electronic Systems, Institute for Basic Science (IBS), Pohang 37673, Republic of Korea}
	\affiliation{Department of Physics, Pohang University of Science and Technology, Pohang 37673, Republic of Korea}

\date{\today}

\begin{abstract}
    In an electronic system with various interactions intertwined, revealing the origin of its many-body ground state is challenging and a direct experimental way to verify the correlated nature of an insulator has been lacking. 
Here we demonstrate a way to unambiguously distinguish a paradigmatic correlated insulator, a Mott insulator, from a trivial band insulator based on their distinct chemical behavior for a surface adsorbate using 1\textit{T}-TaS$_2$, which has been debated between a spin-frustrated Mott insulator or a spin-singlet trivial insulator.
We start from the observation of different sizes of spectral gaps on different surface terminations and show that potassium adatoms on these two surface layers behave in totally different ways.
This can be straightforwardly understood from distinct properties of a Mott and a band insulators due to the fundamental difference of a half and a full-filled orbital involved respectively.  
This work not only solves an outstanding problem in this particularly interesting material but also provides a simple touchstone to identify the correlated ground state of electrons experimentally.
 \end{abstract}

\maketitle


\textit{Introduction.}---When a material has strong interactions in various degrees of freedom, such as charge, spin, orbital, lattice, disorder, and topology, our understanding of the system's ground state and phase evolution is largely limited, and intriguing quantum phenomena emerge.
    The competition of a few different interactions often makes it truly challenging to identify the origin of the material's ground state, such as the high-temperature superconductivity~\cite{Keimer2015}, which would be crucial to control its macroscopic properties and functionalize the material.
    For a simple but paradigmatic example, the Mott insulating state is driven by strong on-site Coulomb interaction of electrons ($U$) for a material with a half-filled electron band, which would otherwise form a metallic state~\cite{Mott1968}. 
    This state is usually distinguished by the existence of an energy gap in spectroscopy against the theoretical prediction of a gapless state without the inclusion of $U$~\cite{Mott1968}. 
    However, there has been no direct experimental way to verify the half-filled insulating state and distinguish it from a band insulator with a full-filled electron band.  
    The situation becomes more complicated when there exist competing insulating states of different origins, such as the charge or spin density wave states. 
    In this Letter, we introduce a simple but powerful method to straightforwardly distinguish a Mott insulator from a band insulator based on the chemical difference in a half and a full-filled electronic orbital.

    As a critical example to show the limitation in distinguishing a Mott insulator under various different interactions, we take a layered transition metal dichalcogenide 1\textit{T}-TaS$_2$.
    This material has been assumed widely as a Mott insulator with a coexisting charge-density-wave state~\cite{Fazekas1979,Giambattista1990,Kim1994,Kim1996,Zwick1998,Pillo2000,Rossnagel2005,Sipos2008,Xu2010,Ang2012,Ang2013,Stojchevska2014,Lahoud2014,Ang2015,Cho2015,Cho2016,Law2017,Cho2017,Ribak2017,Ligges2018,Lutsyk2018,Skolimowski2019,Park2019,Cheng2020} but there exist a few other degrees of freedom to be considered.
    It undergoes a characteristic series of phase transitions from a metal to a commensurate charge-density-wave (CCDW) phase through a nearly commensurate charge-density-wave (NCCDW) phases~\cite{Wilson1975}.
    A common unit cell of CDW phases has 13 Ta atoms distorted into a David-star(DS)-shape cluster to form a $\sqrt{13}\times\sqrt{13}$ superstructure~\cite{Wilson1975}.
    13 5$d$ electrons of a unit cell, one from each Ta atom, form 6 full-filled and one half-filled band with its electron localized strongly on the central Ta atom. 
    The system exhibits a metal-insulator transition accompanying the NCCDW-CCDW transition and the strong localization of the half-filled electron, a very small bandwidth to amplify $U$, provides a natural ground of a Mott insulator model~\cite{Fazekas1979}.
    The model prevailed the scene for more than three decades~\cite{Fazekas1979,Giambattista1990,Kim1994,Kim1996,Zwick1998,Pillo2000,Rossnagel2005,Sipos2008,Xu2010,Ang2012,Ang2013,Stojchevska2014,Lahoud2014,Ang2015,Cho2015,Cho2016,Law2017,Cho2017,Ribak2017,Ligges2018,Lutsyk2018,Skolimowski2019,Park2019,Cheng2020} and fueled the recent proposal of a quantum spin liquid state to explain the absence of a magnetic ordering down to a very low temperature~\cite{Law2017,Ribak2017}.

\begin{figure*}[t]
\includegraphics[width=17.3 cm]{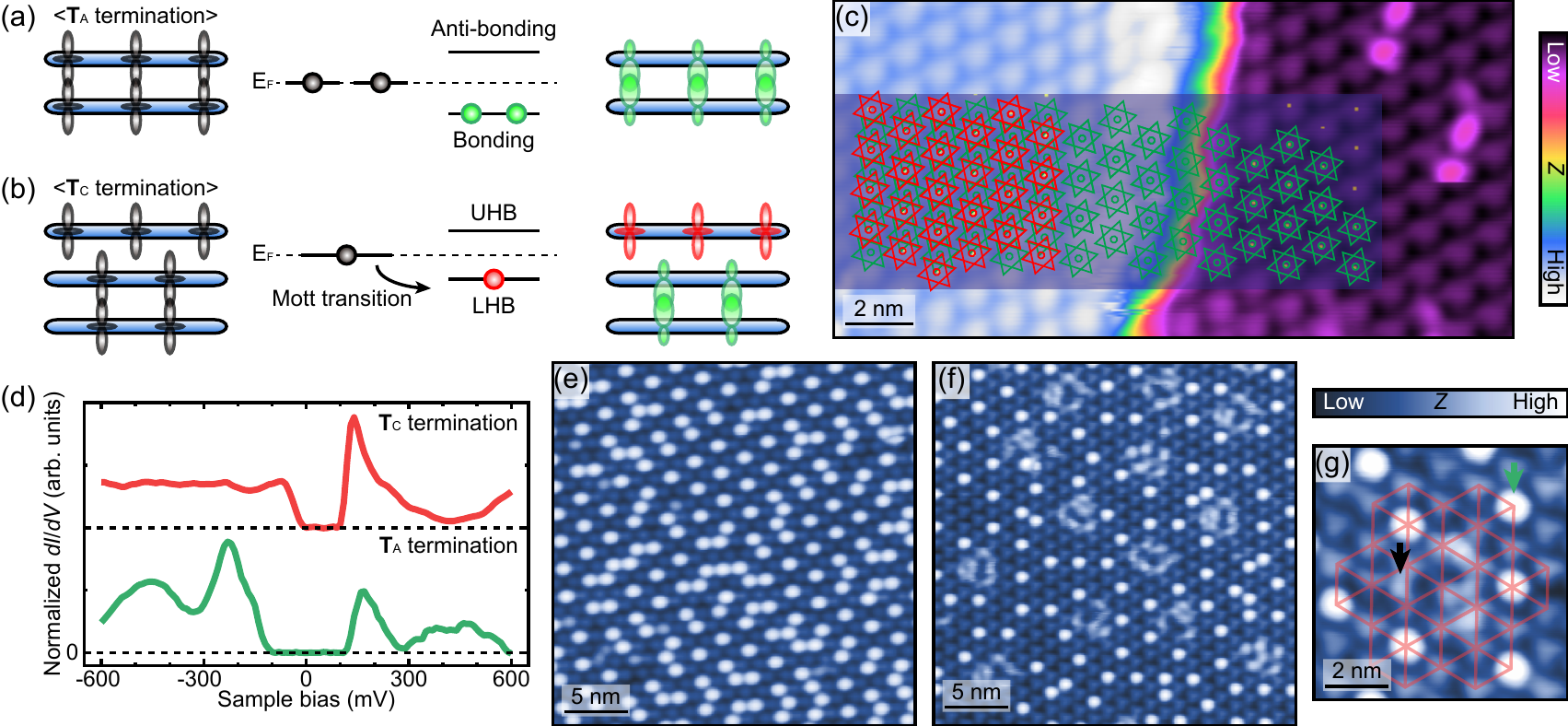}
\caption{\label{stm}
Distinct adsorption behaviors of K atoms on different surface terminations of 1\textit{T}-TaS$_2$.
(a) and (b)~Orbital schematics for inter- (\textbf{T}\textsubscript{A}) and intra- (\textbf{T}\textsubscript{C}) bilayer terminations. Left (right) panels are those without (with) interlayer coupling. Note that only unpaired $d_{z^2}$ orbitals at the center of CDW unit cells are illustrated. Green and red colors represent the paired and unpaired electrons, respectively, after the interlayer coupling.
(c)~Determination of the \textbf{T}\textsubscript{C} stacking of two neighboring surface layers across a single layer step.
Yellow dots overlaid on the image indicate the center of the DS, obtained from the numerical process.
(d)~Normalized $dI/dV$ spectra of neighboring surface layers with different terminations without K adatoms; red (green) spectrum from the \textbf{T}\textsubscript{C} (\textbf{T}\textsubscript{A}) termination in the left (right) part of (c).
(e) and (f)~STM images of \textbf{T}\textsubscript{C}- and \textbf{T}\textsubscript{A}-terminated surfaces with K adatoms. The adatom coverage is 0.38 and 0.33 per DS unit cell in (e) and (f), respectively.
(g)~An enlarged STM image of (f). While all adatoms in (e) are adsorbed at the DS center, they have two adsorption sites in (f) as shown with black (a site between DS unit cells) and green (one on the center) arrows in (g).
}
\end{figure*}

    However, the Mott insulator model is seriously challenged by a few recent band structure calculations~\cite{Ritschel2015,Ritschel2018,Lee2019}.
    Very importantly, the above model ignores the interlayer hopping of electrons, which can crucially affect the bandwidth. 
   The experiments seem to favor a bilayer stacking order, the layers within the bilayer stacked without a lateral translation~\cite{Ishiguro1991,Witte2019,Stahl2020,Butler2020,Wang2020} [a \textbf{T}\textsubscript{A} stacking, Fig.~\ref{stm}(a)] and each bilayer unit stacked with a translation of a half unit cell~\cite{Witte2019,Butler2020} [a \textbf{T}\textsubscript{C} stacking, Fig.~\ref{stm}(b)], where half-filled Ta 5$d_{z^2}$ electrons within bilayers form a spin-singlet state to lead to a trivial bonding-antibonding energy gap against the Mott insulator model. 
    In our point of view, while tracking the intriguing interlayer ordering is very important, this situation manifests the weakness of the current method to verify a Mott insulating state, which relies crucially on the band structure calculation.
    
\textit{Methods.}---1\textit{T}-TaS$_2$ single crystal was cleaved in high vacuum and K atoms were deposited in a ultra high vacuum at room temperature.
	STM images were obtained at 4.4~K by a constant-current mode with a sample bias of -600 (\textbf{T}\textsubscript{C}) or -800~meV (\textbf{T}\textsubscript{A}) depending on the terminations~\cite{suppl}.
	A lock-in technique with a modulation of 1~kHz was utilized for $dI/dV$ measurements.
	After the measurements on the adsorption behavior, K adatoms were removed by through mild electric field applied by an STM tip in order to determine the layer stacking ~\cite{suppl}.

	First-principles calculations were performed using the projected augmented plane-wave method implemented in the Vienna \textit{ab initio} simulation package~\cite{Blochl1994,Kresse1996} and the generalized gradient approximation for the exchange and correlation potential~\cite{Perdew1996}. Spin-orbit coupling was also included. Electronic wave functions were expanded in plane waves with an energy cutoff of 400~eV. The isolated and K-adsorbed TaS$_2$ layers were considered within supercell geometries where the interlayer spacing was 20~{\AA}. Geometries were relaxed until the forces on each atom were less than 0.01~eV~{\AA}$^{-1}$. We used a 7$\times$7$\times$1 and 3$\times$3$\times$1 \textbf{k}-point grid to sample the entire Brillouin zone for the pristine and K-adsorbed 1\textit{T}-TaS$_2$ system, respectively.

\textit{Different surface terminations with distinct adsorption behavior.}---The bilayer stacking would manifest itself as the existence of two distinct, inter- and intra-bilayer, terminations (\textbf{T}\textsubscript{A} and \textbf{T}\textsubscript{C} terminations, respectively) on the cleaved surface [Figs.~\ref{stm}(a) and \ref{stm}(b)]. 
    Indeed, a very recent STM experiment observed both terminations but with different sizes of the band gap~\cite{Butler2020}.
    However, the nature of each band gap is unclear. 
    We pick up two neighboring surface layers separated by a single-layer-height step [Figs.~\ref{stm}(c) and \ref{stm}(d)] and assign accurately the centers of DS clusters through the numerical processing of the STM images and compared them quantitatively between the neighboring layer.
    The top layer corresponds to the \textbf{T}\textsubscript{C} termination, which inevitably requests the bottom one as \textbf{T}\textsubscript{A} based on the alternative \textbf{T}\textsubscript{A}-\textbf{T}\textsubscript{C} stacking order at low temperature~\cite{Witte2019,Butler2020}.
    They both exhibit insulating tunneling spectra [the normalized differential tunneling conductance ($dI/dV$)] but with different band gaps of 250 and 400~meV as shown in Fig.~\ref{stm}(d), which is quantitatively consistent with the recent STM results~\cite{Butler2020,suppl}.

\begin{figure*}[t]
\includegraphics[width=17.7 cm]{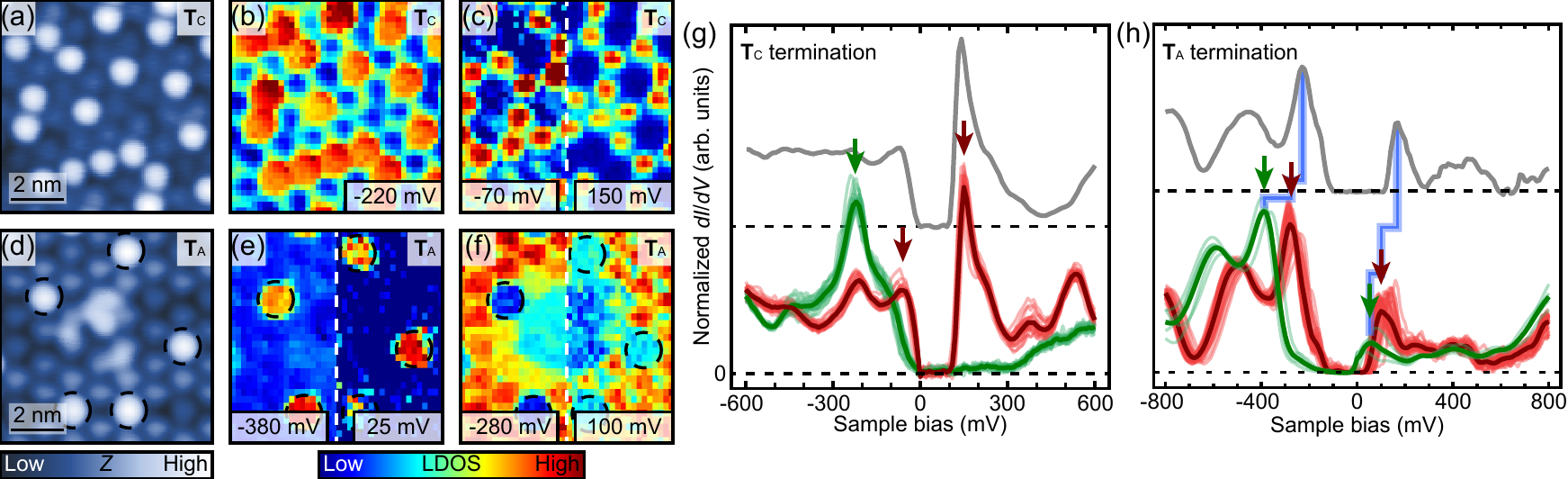}
\caption{\label{sts}
Electronic states of K adatoms on two different layers.
(a)~A STM image of the \textbf{T}\textsubscript{C}-terminated surface.
(b) and (c)~Corresponding LDOS maps at three different biases indicated.
(d)-(f)~Similar images of the \textbf{T}\textsubscript{A}-terminated surface.
(g) and (h)~Normalized $dI/dV$ spectra obtained on the \textbf{T}\textsubscript{C}- and \textbf{T}\textsubscript{A}-terminated surfaces. Red (green) spectra are obtained at the center of the bare (K-adsorbed) CDW unit cells in (a) and (d). Thick and dark spectra are the average of them.
Biases for (b), (c), (e), and (f) are marked with arrows.
The spectra of the pristine sample without adsorbates are plotted in grey for the comparison.
}
\end{figure*}

    These two terminations are sharply contrasted by a simple electron-donating adsorbate of K as illustrated in the STM images of Figs.~\ref{stm}(e)-\ref{stm}(g).
    The most apparent is that the adsorption behaviors of K atoms are distinct on different terminations.
    On the \textbf{T}\textsubscript{C} termination [Fig.~\ref{stm}(e)], all K atoms sit on the center of a DS cluster as reported recently~\cite{Lee2020}, and the surface unit cells bifurcate into two groups, DS with and without K adatom at their center.
    On the other hand, we observe two different adsorbate configurations on the \textbf{T}\textsubscript{A} termination.
    The bright ones are at the center of the DS clusters as on the \textbf{T}\textsubscript{C} termination, but there exist off-centered adsorbates [medium contrast protrusions in Figs.~\ref{stm}(f) and \ref{stm}(g)] mostly between neighboring DS clusters.
    The off-centered adsorbates are metastable configurations, which are prone to hop and easily perturbed by STM tips.
    The distinct adsorption behavior of the neighboring terraces is uniformly observed over different areas surfaces and in different cleavages.

\textit{Distinct electronic states of different surface terminations with adsorbates.}---The difference between two different terminations becomes more evident in the spectroscopic measurements. 
The DS clusters with central K adsorbates exhibit strong local density of states (LDOS) at different energies, -220~meV on the \textbf{T}\textsubscript{C} [Figs.~\ref{sts}(a) and \ref{sts}(b)], and 25 and -380~meV on the \textbf{T}\textsubscript{A} termination [Figs.~\ref{sts}(d) and \ref{sts}(e)] as shown in the LDOS maps.
The corresponding spectra reveal more important details.
The DS clusters without K on the \textbf{T}\textsubscript{C} termination show no change of the LDOS while those with K have a single strong LDOS peak at -220~meV replacing the two peaks defining the energy gap at the Fermi level [Fig.~\ref{sts}(g)]~\cite{suppl}. 
This is natural for a Mott insulator since the half-filled electron in a unit cell with a K adsorbate would be fully saturated by the electron from the adsorbate.
Since the full-filled electron orbital is strongly localized, the neighboring bare unit cells would have little doping effect [Figs.~\ref{sts}(c) and \ref{sts}(g)].
In stark contrast, on the \textbf{T}\textsubscript{A} termination, the two LDOS peaks defining the band gap are rigidly shifted to lower energies for the DS clusters without an adsorbate and to even lower energy for those with K [Fig.~\ref{sts}(h)]~\cite{suppl}.
This straightforwardly indicates that the K adsorbates dope the system. 
The rigid shift upon doping is what is expected in a band insulator.
The doping effect is stronger on the DS cluster with K adsorbate and weaker on those without [see their LDOS maps in Figs.~\ref{sts}(e) and \ref{sts}(f)].

\begin{figure*}[t]
\includegraphics[width=16.7 cm]{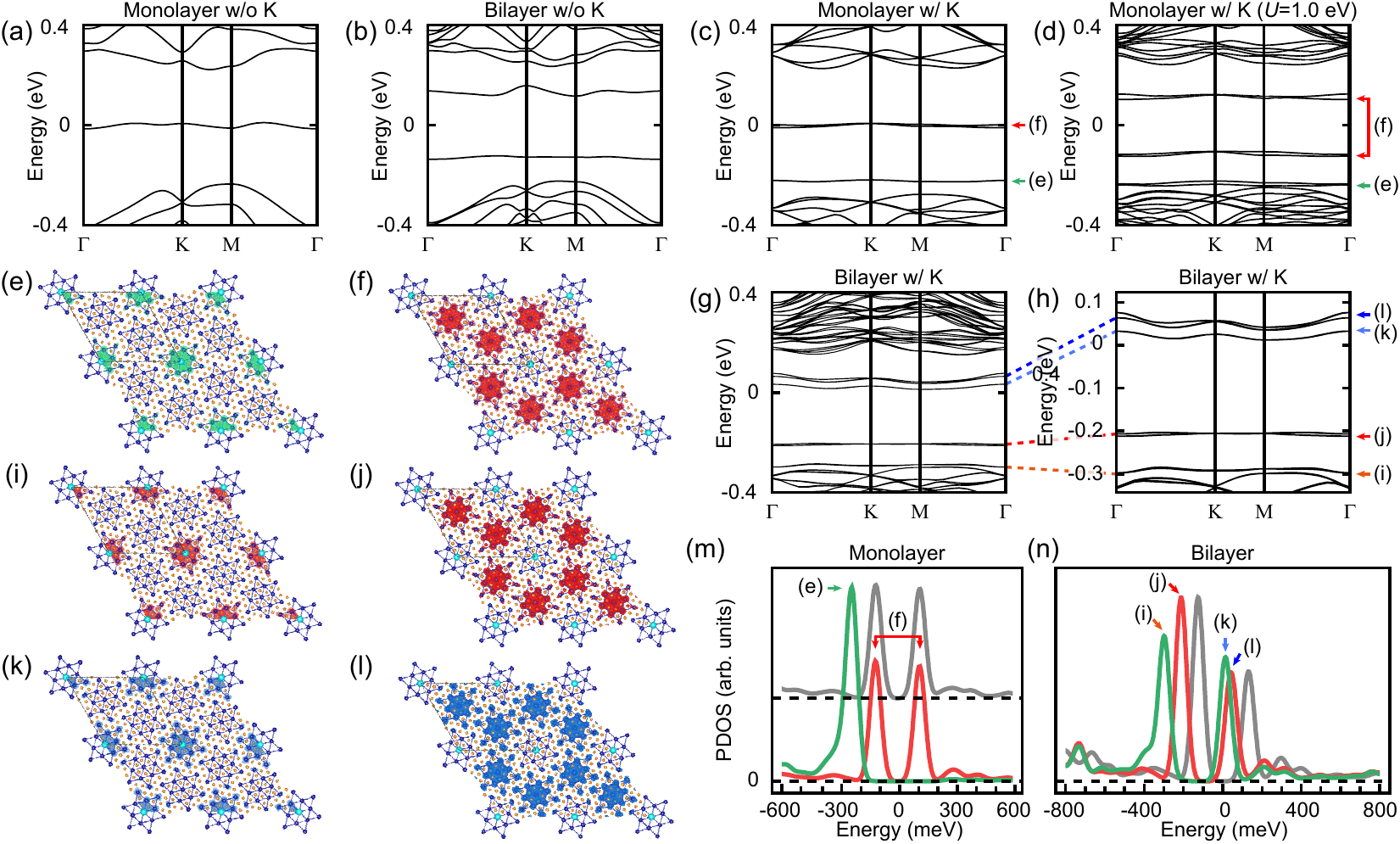}
\caption{\label{dft}
Electronic band structures and local density of states calculated for K-adsorbed 1{\textit {T}}-TaS$_2$.
(a) and (b)~Band structures of the monolayer and bilayer (\textbf{T}\textsubscript{A}-stacked) 1{\textit {T}}-TaS$_2$, respectively.
(c) and (d)~Band structures of a monolayer with K adatoms in a $(\sqrt{3}\times\sqrt{3})R30^{\circ}$ superstructure without and with the on-site Coulomb repulsion $U$, respectively.
(e) and (f)~Spatial distributions of electrons within the bands marked in (c) and (d).
(g)~A band structure of a bilayer 1{\textit {T}}-TaS$_2$ in the \textbf{T}\textsubscript{A} stacking with K adatoms in the same superstructure as in (c).
(h)~An enlargement of (g) near the Fermi level.
(i)-(l)~Electron spatial distributions of the bands marked in (h).
(m)~5$d_{z^2}$ partial density of states for the K adsorption on the monolayer, and (n) on the bilayer 1\textit{T}-TaS$_2$. Gray, red, and green plots represent the PDOS in the pristine layer, and the bare and the K-adsorbed DS unit cells in the K-adsorbed layer. Each of the PDOS peaks corresponds to bands indicated in (d) or (h) whose spatial distributions are shown in (e), (f), or (i)-(l). Blue, orange, and cyan atoms in (e), (f), and (i)-(l) represent Ta, S, and K, respectively.
See the Supplemental Material~\cite{suppl} for the side views of the charge distribution.
}
\end{figure*}

\textit{DFT calculations with a single and a bilayer model.}---All experimental results are well explained by the DFT calculations with an isolated single and a bilayer layer model for the \textbf{T}\textsubscript{C} and \textbf{T}\textsubscript{A} termination, respectively. 
    The validity of the single-layer model can be confirmed by a trilayer model~\cite{suppl} and is easily understood from a marginal overlap of half-filled $d_{z^2}$ electrons between the top and the second layer.
    Even before any calculation, the difference of the electronic system is obvious.
    In contrast to the single layer with half-filled electrons, there exist an even number of $d_{z^2}$ electrons in a double-layer unit cell, which would form an interlayer bonding on each unit cell to fall into a simple band insulator [Fig.~\ref{stm}(a)].
    These result in totally different single-particle band structures, a metallic band at the Fermi level [Fig.~\ref{dft}(a)] and insulating bonding-antibonding bands [Fig.~\ref{dft}(b)] in a single and a double layer, respectively.  
    As well known, introducing electron correlation $U$ in the single layer would open a Mott gap but does not affect substantially the bonding-antibonding bands~\cite{suppl}.
    
    The K adsorption on both terminations is modeled with a periodic occupation of 1/3 of DS clusters, which results in a  $(\sqrt{3}\times\sqrt{3})R30^{\circ}$ superstructure. 
    As detailed previously, on the \textbf{T}\textsubscript{C} termination, this superstructure is highly favored due to the dipole-dipole interaction of adsorbates~\cite{Lee2020}. 
    When K atoms are deposited, its 4$s$ electron forms a strong local bonding with $d_{z^2}$ electron at the center of the DS [Fig.~\ref{dft}(e)] and a full-filled band at -200 meV [Fig.~\ref{dft}(c)]. 
    The electrons in the bare DS clusters are almost intact [Fig.~\ref{dft}(f)] with their partially-filled band remaining at the Fermi level [Fig.~\ref{dft}(c)] or Mott Hubbard bands [Fig.~\ref{dft}(d)] when $U$ is excluded or included, respectively. 
    The corresponding partial density of states [PDOS, Fig.~\ref{dft}(m)] reproduce well the spectroscopic data~\cite{suppl}, two distinct spectra on a K-adsorbed and a bare DS cluster, as mentioned above [Fig.~\ref{sts}(g)].
    On the \textbf{T}\textsubscript{A} termination, in clear contrast, the 4$s$ electron is donated to the $d_{z^2}$ bands to shift them rigidly.
    There exist different amounts of shifts for the DS clusters with and without K adsorbates, a larger shift for the former, as observed in the experiment, giving rise to basically two sets of bands with wave functions on different unit cells [Figs.~\ref{dft}(g), \ref{dft}(h), and \ref{dft}(i)-\ref{dft}(l)].  
    The inclusion of $U$ does not affect the overall band structure but only increases the band gap marginally ($\sim$15~meV)~\cite{suppl}.
    The PDOS changes [Fig.~\ref{dft}(n)] match excellently with the experimental data [Fig.~\ref{sts}(h)]~\cite{suppl}.

\begin{figure}[t]
\includegraphics[width=8.3 cm]{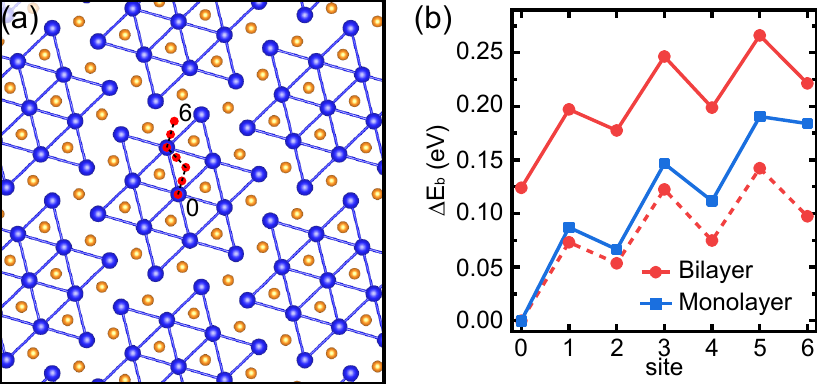}
\caption{\label{binding}
Adsorption energies of K adatoms on different terminations.
(a)~Representative adsorption sites of K, and (b)~their relative binding energies ($\Delta E_b$) to the most stable site of a monolayer (blue). Those on a \textbf{T}\textsubscript{A}-stacked bilayer (red) are shifted for the comparison (dashed red line).
}
\end{figure}

\textit{Details of the adsorption behavior.}---As we mentioned above, the difference of the surface termination appears not only in spectroscopy but also in the adsorbate distribution as the existence of the minor adsorption sites on the \textbf{T}\textsubscript{A} termination. 
    This can also be explained within the DFT calculations. 
    In principle, there are 28 symmetrically inequivalent adsorption sites in a DS cluster~\cite{Lee2020}.  
    We first exclude those on top of surface sulfur atoms, which are energetically unfavored.
     The adsorption energy is found to be minimized at the DS center but depends on the termination as -2.85 and -2.73~eV on \textbf{T}\textsubscript{C} and \textbf{T}\textsubscript{A} termination, respectively. 
     A larger adsorption energy gain is expected on the \textbf{T}\textsubscript{C} termination due to the localized half-filled electrons as discussed above. 
     As shown for 7 representative sites [Fig.~\ref{binding}(a)] with different distances from the center, the adsorption energy increases rather linearly with the distance from the center [Fig.~\ref{binding}(b)].
     We note the relatively large difference of the adsorption energy between the site 5 and 6 on the \textbf{T}\textsubscript{A} than the \textbf{T}\textsubscript{C} termination.
     Note also that the energy difference between the site 6 and the most stable site (site 0) is much large on the \textbf{T}\textsubscript{C} termination.
     Thus, the adatom migration to a more stable sites toward the center of a DS cluster is much easier on the \textbf{T}\textsubscript{C} termination.
     These adsorption-energy landscapes qualitatively explain the absence/presence of adsorbates on the site 6 in the experiment on the \textbf{T}\textsubscript{C}/\textbf{T}\textsubscript{A} termination~\cite{suppl}.
    While the difference in the adsorbate distribution is easily observable in the topographic imaging, its origin is not as fundamental as the spectroscopic difference on different terminations.
    However, roughly speaking, the origin of the difference can be traced back to the stronger bonding of the electron-donating adsorbate on the Mott insulator due to the half-filled electron.
   Combining all the experimental and theoretical result consistently, we introduce that the adsorption of monovalent adsorbates can be a direct and unique touchstone to distinguish a Mott insulator from a band insulator or other types of insulators based on filled orbitals.

\bigskip

This work was supported by the Institute for Basic Science (Gradnt No. IBS-R014-D1). K.-H.~Jin is supported by the Institute for Basic Science (Grant No. IBS-R014-Y1).


\begin{thebibliography}{99}
\bibliographystyle{apsrev4-1}
\setcitestyle{square}

\bibitem{Keimer2015}
{B. Keimer, S. A. Kivelson, M. R. Norman, S. Uchida, and J. Zaanen, \textit{From quantum matter to high-temperature superconductivity in copper oxides,} Nature \textbf{518}, 179 (2015).}

\bibitem{Mott1968}
{N. F. Mott, \textit{Metal-Insulator Transition,} Rev. Mod. Phys. \textbf{40}, 677 (1968).}



\bibitem{Fazekas1979}
{P. Fazekas and E. Tosatti, \textit{Electrical, structural and magnetic properties of pure and doped $\mathrm{1T-TaS}_{2}$,} Philos. Mag. B \textbf{39}, 229 (1979).}

\bibitem{Giambattista1990}
{B. Giambattista, C. G. Slough, W. W. McNairy, and R. V. Coleman, \textit{Scanning tunneling microscopy of atoms and charge-density waves in ${\mathrm{1}}T{\mathrm{-TaS}}_{2}$, ${\mathrm{1}}T{\mathrm{-TaSe}}_{2}$, and ${\mathrm{1}}T{\mathrm{-VSe}}_{2}$,} Phys. Rev. B \textbf{41}, 10082 (1990).}

\bibitem{Kim1994}
{J.-J. Kim, W. Yamaguchi, T. Hasegawa, and K. Kitazawa, \textit{Observation of Mott Localization Gap Using Low Temperature Scanning Tunneling Spectroscopy in Commensurate ${\mathrm{1}}T{\mathrm{-TaS}}_{2}$,} {Phys. Rev. Lett.} \textbf{73}, 2103 (1994).}

\bibitem{Kim1996}
{J.-J. Kim, I. Ekvall, and H. Olin, \textit{Temperature-dependent scanning tunneling spectroscopy of ${\mathrm{1}}T{\mathrm{-TaS}}_{2}$,} {Phys. Rev. B} \textbf{54}, 2244 (1996).}

\bibitem{Zwick1998}
{F. F.Zwick \textit{et al.}, \textit{Spectral Consequences of Broken Phase Coherence in ${\mathrm{1}}T{\mathrm{-TaS}}_{2}$,} {Phys. Rev. Lett.} \textbf{81}, 1058 (1998).}

\bibitem{Pillo2000}
{T. Pillo \textit{et al.}, \textit{Interplay between electron-electron interaction and electron-phonon coupling near the Fermi surface of $1T\ensuremath{-}{\mathrm{TaS}}_{2}$,} {Phys. Rev. B} \textbf{62}, 4277 (2000).}

\bibitem{Rossnagel2005}
{K. Rossnagel, E. Rotenberg, H. Koh, N. V. Smith, and L. Kipp, \textit{Continuous Tuning of Electronic Correlations by Alkali Adsorption on Layered $1T\mathrm{\text{\ensuremath{-}}}{\mathrm{TaS}}_{2}$,} {Phys. Rev. Lett.} \textbf{95}, 126403 (2005).}

\bibitem{Sipos2008}
{B. Sipos \textit{et al.}, \textit{From Mott state to superconductivity in ${\mathrm{1}}T{\mathrm{-TaS}}_{2}$,} {Nat. Mater.} \textbf{7}, 960 (2008).}

\bibitem{Xu2010}
{P. Xu \textit{et al.}, \textit{Superconducting phase in the layered dichalcogenide ${\mathrm{1}}T{\mathrm{-TaS}}_{2}$ upon inhibition of the metal-insulator transition,} {Phys. Rev. B} \textbf{81}, 172503 (2010).}

\bibitem{Ang2012}
{R. Ang \textit{et al.}, \textit{Real-Space Coexistence of the Melted Mott State and Superconductivity in Fe-Substituted $1T\mathrm{\text{\ensuremath{-}}}{\mathrm{TaS}}_{2}$,} {Phys. Rev. Lett.} \textbf{109}, 176403 (2012).}

\bibitem{Ang2013}
{R. Ang \textit{et al.}, \textit{Superconductivity and bandwidth-controlled {Mott} metal-insulator transition in 1$T$-TaS${}_{2\ensuremath{-}x}$Se${}_{x}$,} {Phys. Rev. B} \textbf{88}, 115145 (2013).}

\bibitem{Stojchevska2014}
{L. Stojchevska \textit{et al.}, \textit{Ultrafast Switching to a Stable Hidden Quantum State in an Electronic Crystal,} {Science} \textbf{344}, 177 (2014).}

\bibitem{Lahoud2014}
{E. Lahoud, O. N. Meetei, K. B. Chaska, A. Kanigel, and N. Trivedi, \textit{Emergence of a Novel Pseudogap Metallic State in a Disordered 2D Mott Insulator,} {Phys. Rev. Lett.} \textbf{112}, 206402 (2014).}

\bibitem{Ang2015}
{R. Ang \textit{et al.}, \textit{Atomistic origin of an ordered superstructure induced superconductivity in layered chalcogenides,} {Nat. Commun.} \textbf{6}, 6091 (2015).}

\bibitem{Cho2015}
{D. Cho, Y.-H. Cho, S.-W. Cheong, K.-S. Kim, and H. W. Yeom, \textit{Interplay of electron-electron and electron-phonon interactions in the low-temperature phase of ${\mathrm{1}}T{\mathrm{-TaS}}_{2}$,} {Phys. Rev. B} \textbf{92}, 085132 (2015).}

\bibitem{Cho2016}
{D. Cho \textit{et al.}, \textit{Nanoscale manipulation of the Mott insulating state coupled to charge order in ${\mathrm{1}}T{\mathrm{-TaS}}_{2}$,} {Nat. Commun.} \textbf{7}, 10453 (2016).}

\bibitem{Law2017}
{K. T. Law and P. A. Lee, \textit{${\mathrm{1}}T{\mathrm{-TaS}}_{2}$ as a quantum spin liquid,} {Proc. Natl. Acad. Sci.} \textbf{114}, 6996 (2017).}

\bibitem{Cho2017}
{D. Cho \textit{et al.}, \textit{Correlated electronic states at domain walls of a Mott-charge-density-wave insulator ${\mathrm{1}}T{\mathrm{-TaS}}_{2}$,} {Nat. Commun.} \textbf{8}, 392 (2017).}

\bibitem{Ribak2017}
{A. Ribak \textit{et al.}, \textit{Gapless excitations in the ground state of ${\mathrm{1}}T{\mathrm{-TaS}}_{2}$,} {Phys. Rev. B} \textbf{96}, 195131 (2017).}

\bibitem{Ligges2018}
{M. Ligges \textit{et al.}, \textit{Ultrafast Doublon Dynamics in Photoexcited $1T$-${\mathrm{TaS}}_{2}$,} {Phys. Rev. Lett.} \textbf{120}, 166401 (2018).}

\bibitem{Lutsyk2018}
{I. Lutsky \textit{et al.}, \textit{Electronic structure of commensurate, nearly commensurate, and incommensurate phases of $1T\ensuremath{-}\mathrm{Ta}{\mathrm{S}}_{2}$ by angle-resolved photoelectron spectroscopy, scanning tunneling spectroscopy, and density functional theory,} {Phys. Rev. B} \textbf{98}, 195425 (2018).}

\bibitem{Skolimowski2019}
{J. Skolimowski, Y. Gerasimenko, and R. \ifmmode \check{Z}\else \v{Z}\fi{}itko, \textit{Mottness collapse without metallization in the domain wall of the triangular-lattice Mott insulator $1T\ensuremath{-}{\mathrm{TaS}}_{2}$,} {Phys. Rev. Lett.} \textbf{122}, 036802 (2019).}

\bibitem{Park2019}
{J. W. Park, G. Y. Cho, J. Lee, and H. W. Yeom, \textit{Emergent honeycomb network of topological excitations in correlated charge density wave,} {Nat. Commun.} \textbf{10}, 4038 (2019).}

\bibitem{Cheng2020}
{L. Cheng \textit{et al.}, \textit{Renormalization of the Mott gap by lattice entropy: The case of 1T-${\mathrm{TaS}}_{2}$,} {Phys. Rev. Res.} \textbf{2}, 023064 (2020).}




\bibitem{Wilson1975}
{J. A. Wilson, F. J. Di Salvo, and S. Mahajan, \textit{Charge-density waves and superlattices in the metallic layered transition metal dichalcogenides,} {Adv. Phys.} \textbf{24}, 117 (1975).}






\bibitem{Ritschel2015}
{T. Ritschel \textit{et al.}, \textit{Orbital textures and charge density waves in transition metal dichalcogenides,} {Nat. Phys.} \textbf{11}, 328 (2015).}

\bibitem{Ritschel2018}
{T. Ritschel, H. Berger, and J. Geck, \textit{Stacking-driven gap formation in layered 1T-${\mathrm{TaS}}_{2}$,} {Phys. Rev. B} \textbf{98}, 195134 (2018).}

\bibitem{Lee2019}
{S.-H. Lee, J. S. Goh, and D. Cho, \textit{Origin of the Insulating Phase and First-Order Metal-Insulator Transition in $1T\text{\ensuremath{-}}{\mathrm{TaS}}_{2}$,} {Phys. Rev. Lett.} \textbf{122}, 106404 (2019).}

\bibitem{Wang2020}
{Y. D. Wang \textit{et al.}, \textit{Band insulator to Mott insulator transition in 1\textit{T}-TaS$_2$,} {Nat. Commun.} \textbf{11}, 4215 (2020).}

\bibitem{Butler2020}
{C. J. Butler, M. Yoshida, T. Hanaguri, and Y. Iwasa, \textit{Mottness versus unit-cell doubling as the driver of the insulating state in ${\mathrm{1}}T{\mathrm{-TaS}}_{2}$,} {Nat. Commun.} \textbf{11}, 2477 (2020).}







\bibitem{Ishiguro1991}
{T. Ishiguro and H. Sato, \textit{Electron microscopy of phase transformations in 1T-${\mathrm{TaS}}_{2}$,} {Phys. Rev. B} \textbf{44}, 2046 (1991).}

\bibitem{Witte2019}
{G. von Witte \textit{et al.}, \textit{Surface structure and stacking of the commensurate $(\sqrt{13}\ifmmode\times\else\texttimes\fi{}\sqrt{13})R13.{9}^{\ensuremath{\circ}}$ charge density wave phase of $1T\ensuremath{-}{\mathrm{TaS}}_{2}(0001)$,} {Phys. Rev. B} \textbf{100}, 155407 (2019).}

\bibitem{Stahl2020}
{W. Stahl \textit{et al.}, \textit{Collapse of layer dimerization in the photo-induced hidden state of {1T-TaS$_2$},} {Nat. Commun.} \textbf{11}, 1247 (2020).}



\bibitem{suppl}
{See the Supplemental Material at [URL will be given by the publisher] for the details, which includes Refs.~\cite{Cho2016,Ma2016,Wang2020,Butler2020}.}


\bibitem{Blochl1994}
{P. E. Bl\"ochl, \textit{Projector augmented-wave method,} {Phys. Rev. B} \textbf{50}, 17953 (1994).}

\bibitem{Kresse1996}
{G. Kresse, and J. Furthm\"uller, \textit{Efficient iterative schemes for ab initio total-energy calculations using a plane-wave basis set,}  {Phys. Rev. B} \textbf{54}, 11169 (1996).}

\bibitem{Perdew1996}
{J. P. Perdew, K. Burke, and M. Ernzerhof, \textit{Generalized Gradient Approximation Made Simple,} {Phys. Rev. Lett.} \textbf{77}, 3865 (1996).}


\bibitem{Lee2020}
{J. Lee \textit{et al.}, \textit{Honeycomb-Lattice Mott Insulator on Tantalum Disulphide,} {Phys. Rev. Lett.} \textbf{125}, 096403 (2020).}







\bibitem{Ma2016}
{L. Ma \textit{et al.}, \textit{A metallic mosaic phase and the origin of Mott-insulating state in 1T-TaS$_2$,} {Nat. Commun.} \textbf{7}, 10956 (2016).}








\end{thebibliography}
\end{document}